# Development and prospective validation of a prostate cancer detection, grading, and workflow optimization system at an academic medical center

Ramin Nateghi[1], Ruoji Zhou[2], Madeline Saft[1], Marina Schnauss[1], Clayton Neill[1], Ridwan Alam[1], Nicole Handa[1], Mitchell Huang[1], Eric V Li[1], Jeffery A Goldstein[2], Hiten D Patel[1], Edward M Schaeffer[1], Menatalla Nadim[2], Fattaneh Pourakpour[2], Bogdan Isaila[2], Christopher Felicelli[2], Vikas Mehta[2], Behtash G Nezami[2], Ashley E Ross[1], Ximing J Yang[2], Lee AD Cooper[2,3,4]

1. Department of Urology, Northwestern University Feinberg School of Medicine, Chicago, IL USA
2. Department of Pathology, Northwestern University Feinberg School of Medicine, Chicago, IL USA
3. Center for Computational Imaging and Signal Analytics, Northwestern University Feinberg School of Medicine, Chicago, IL USA
4. Chan Zuckerberg Biohub, Chicago IL USA

## Summary

**Background** Artificial intelligence may assist healthcare systems in meeting increasing demand for pathology services while maintaining diagnostic quality and reducing turnaround time and costs. We aimed to investigate the performance of an institutionally developed system for prostate cancer detection, grading, and workflow optimization and to contrast this with commercial alternatives.

**Methods** From August 2021 to March 2023, we scanned 21,396 slides from 1,147 patients receiving prostate biopsy found to have at least one positive core. We developed models for cancer detection, grading, and screening of equivocal cases for immunohistochemistry ordering. We compared the performance of task-specific prostate models with general-purpose foundation models in a prospectively collected dataset that reflects our patient population. We also evaluated the contributions of a bespoke model designed to improve sensitivity to small cancer foci and perception of low-resolution patterns.

**Findings** We found high concordance with pathologist ground-truth in detection (area under curve 98.5%, sensitivity 95.0%, and specificity 97.8%), ISUP grading (Cohen's kappa 0.869), grade group 3 or higher classification (area under curve 97.5%, sensitivity 94.9%, specificity 96.6%). Screening models could correctly classify 55% of biopsy blocks where immunohistochemistry was ordered with a 1.4% error rate. No statistically significant differences were observed between task-specific and foundation models in cancer detection, although the task-specific model is significantly smaller and faster. Bespoke model performance was superior to alternatives in all tasks. The consensus review confirmed low-grade cancers detected by the model, but these discrepancies would not have changed clinical management of these patients.

**Interpretation** Institutions like academic medical centers that have high scanning volumes and report abstraction capabilities can develop highly accurate computational pathology models for internal use. These models have the potential to aid in quality control role and to improve resource allocation and workflow in the pathology lab to help meet future challenges in prostate cancer diagnosis.

# Introduction

Prostate cancer is the second most diagnosed malignancy in men worldwide, with over 1 million new cases diagnosed per year [1]. While biomarkers and advanced imaging such as Prostate-Specific Antigen (PSA) and multiparametric Magnetic Resonance Imaging (mpMRI) have aided in the detection and risk stratification of prostate cancer, management of this disease ultimately relies on pathologic diagnosis from biopsied tissues. With an aging population worldwide, the volume of prostate biopsy specimens is expected to increase dramatically [2, 3]. Meeting this demand will increase workload of pathologists who are also facing workforce challenges [4]. Artificial intelligence offers a solution to meet demand while maintaining or improving quality through efficiency gains or upskilling general pathologists [5, 6]. Notably, subspecialists typically perform better than general pathologists in prostate cancer detection and grading [7-9], with general pathologists achieving agreement in less than 60% of cases while 90% of disagreements leading to upgrades [13]. AI tools could help bridge this performance where whole-slide imaging (WSI) is used, and reduce costs and delays associated with consultation and immunohistochemical analysis.

Immunohistochemistry (IHC) is an important part of the pathology workflow in prostate cancer with significant implications for costs and delays. IHC triple stain, which comprises AMACR(P504S), p63, and high molecular weight cytokeratin, is mainly ordered to distinguish invasive carcinoma from high-grade prostatic intraepithelial neoplasia and other benign mimickers following initial review of the H&E slides. Ordering IHC introduces delays in diagnosis and creates additional work for pathologists. The cost of IHC analysis can be substantial for payers and the health care system, with a Medicare price of $172.76 and with approximately 1 million biopsies performed annually in the United States [10, 11]. One study found that community pathologists are more than twice as likely than academic pathologists to order IHC (26% vs. 11%) [12], adding costs and delays to patients diagnosed in community-based practices. Few studies have analyzed the impact of AI tools on IHC use and impact to costs and delays in prostate cancer [13].

Multiple commercial tools intended for clinical use have been developed for prostate cancer to aid detection and grading of biopsies [14, 15]. Performance of the FDA-approved Paige prostate cancer detection system has been described in several studies [5, 16, 17]. The Paige system was developed with a weakly-supervised method that learns from slide-level labels to avoid the need for image annotation. IBEX Galen Prostate is a CE-IVD approved tool for prostate cancer detection and grading that was developed using 1.3M annotations of high-power fields to train field-level classifiers [6]. The performance metrics for standalone detection and grading performed by these tools are presented in **Tables S1-2**. While these tools can be used in a standalone second read capacity for quality control, they are intended to aid in primary diagnosis, and so studies also report performance with interactive use, including improvements in tool-assisted read time. These studies emphasize external validation since commercial tools will encounter significant preanalytical variability in deployment.

Prostate cancer detection and grading of biopsies has been a popular research topic. Two papers published in 2020 report similar performance in detection and grading, both using image annotations to develop field-level classifiers and image segmentation models (see **Tables S1-2**) [18, 19]. Data used in these studies supported the PANDA challenge that reported on the performance of models from academic competitors on a large, curated dataset [20]. Internal and external test data used in [18-20] include data from a population-based screening study [19] and were reviewed by multiple uropathologists to achieve accurate ground truth [18, 20]. Other models for prostate cancer detection and grading have been reported, developed and evaluated on PANDA and other data sources [21-25]. More recently, foundation models trained on datasets of hundreds of thousands or millions of WSI have demonstrated strong performance in many tasks including prostate cancer detection and grading [26, 27].

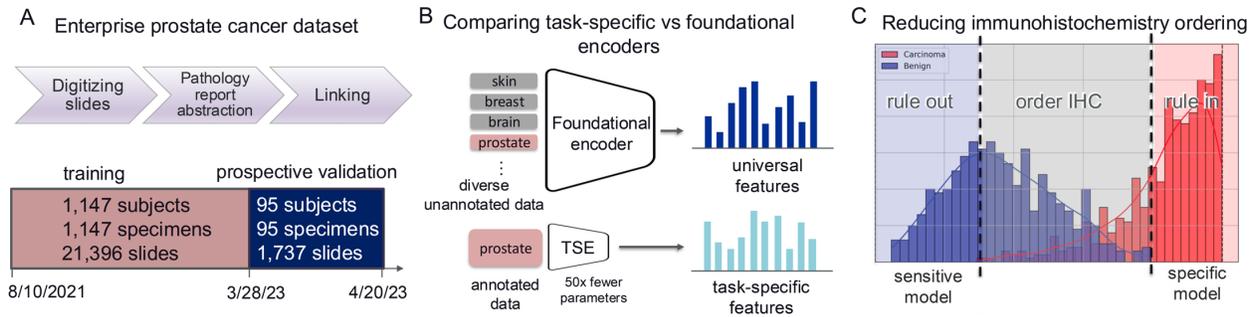

**Figure 1. Development and validation of an institutional AI system for prostate biopsy analysis**. (A) We developed an enterprise-scale prostate cancer dataset over several years to support model development and validation. (B) Using annotated prostate data, we developed a task-specific encoder for prostate cancer detection and grading and compared its capabilities with UNI foundational encoder. (C) Reduction of immunohistochemistry (IHC) ordering, which is often associated with delays, higher costs, and additional workload, such as the re-review of H&E slides upon receiving IHC results.

Data from challenges and academic studies are an invaluable community resource. The selection and exclusion of cases for practical reasons or to balance dataset characteristics is often necessary in building such datasets but may distort expectations of model performance in clinical use. Adjusting the proportions of benign and malignant cases or grades can improve evaluation on less common patterns like high-grade cancers but can also increase the fraction of cases where AI models typically excel. Exclusion based on discordant pathology reviews, imaging or tissue processing artifacts, or selecting for slides with higher cancer extent can also remove more challenging cases from evaluation that would diminish model sensitivity or produce false alarms that require attention.

In this study we describe the development and validation of an institutional AI system for prostate cancer detection and grading and minimization of IHC ordering. This system was developed and validated using data collected at Northwestern Memorial Hospital over multiple years, including a prospectively collected validation dataset that represents our patient population. We show that this model has comparable performance to published data on commercial systems and can reduce immunohistochemical ordering for equivocal biopsy blocks by 55% with an error rate of 1.4%. We further compare foundation models to more efficient prostate-specific models, finding equivalent performance with significant cost savings, and demonstrate the utility of a bespoke model designed to improve detection, grading, and IHC screening tasks.

## Methods

### Datasets

Data for model development and validation were generated at Northwestern Memorial Hospital (NMH) between August 2021 and May 2023 (IRB protocol 00213676). All slide scanning was performed using a Leica Systems Aperio GT 450 scanner with 40x objective magnification. In the period from August 2021 to April 2023 we scanned 23,133 slides from 1,242 patients with at least one positive core for model development purposes (see **Figure 1A**). To collect a representative sample of our patient population for validation, the criteria was changed to all slides from all prostate biopsies for the month of April 2023 for a total of 1,737 slides from 95 subjects. Patients were stratified into internal training, validation, and testing sets (See **Figure S1**). Prior to scanning, slides were wiped to remove any pen markings present on the coverslips. Pathology report data was abstracted from the Northwestern Medicine Electronic Data Warehouse (NM EDW) and normalized by the enterprise data warehouse team using a combination of database query and natural language processing to obtain primary and secondary Gleason patterns for each block and other diagnostic information. A workflow was developed to link scanned slides to the structured pathology and clinical data tables using a combination of barcode reading and optical character recognition with manual review of slide label images. This process is summarized in **Figure S2**. A summary of all slides, blocks, specimens, and subjects is presented in **Table S3**.

For IHC model development, we utilized our entire prostate cancer cohort, excluding cases with IHC staining. The IHC-requested slides scanned from August 2021 to July 2024 were reserved exclusively for testing (See **Figure S1**). We identified slides from blocks where reported cancer extent was less than 20% and with Gleason grade groups of 1 or 2 (these criteria were used to exclude blocks where IHC was ordered for medicolegal reasons and not as a diagnostic aid). The remaining cohort was partitioned into training and validation sets at a 0.9:0.1 ratio, with excluded cases reserved for model testing. This strategy ensured evaluation on challenging cases that reflect real-world low-volume cancer patients typically requiring additional IHC testing.

**Histology encoders and prediction model architecture**

PANDA data was used to develop a task-specific encoder (TSE) for embedding high-power fields (**Figure 1B**). The PANDA dataset comprises 10,616 fully annotated WSIs retrospectively collected from patients undergoing prostate biopsy due to suspected cancer, as summarized in Supplementary **Table S4**. We observed a significant difference in the accuracy of image annotations between the Radboud Medical Center and Karolinska institute datasets, with the annotations from Karolinska lacking spatial precision. We utilized only Radboud slides in our experiments, removing 312 of these slides due to missing annotations or the presence of tissue processing artifacts.

An EfficientNetB0 convolutional network was trained to classify high-power fields using this data (**Figure S3A**). After applying ICC correction **(Figure S4)**, the data was split at the slide level into 80% for training and 20% for in-training validation. A total of 1.41 million 256x256 pixel high-power fields were extracted at 20X objective magnification. Field-level annotations were derived from the provided pixel mask annotations to calculate the proportions of carcinoma and stroma in each tile. Tiles containing at least 25% carcinoma were labeled as carcinoma, while tiles that were entirely benign were labeled as benign. Following training, the network was truncated by removing the last dense classification layers to create a task-specific encoder. The encoder was applied to the fields from the NMH cohort using a nonoverlapping mosaic tiling of tissue regions to generate an embedding for each tile. Encoded tiles from the NMH slides were aggregated by slide and then block to develop and validate detection and grading models. We also evaluate the general-purpose UNI encoder as an alternative to the task-specific encoder trained with PANDA. UNI was not fine-tuned and was applied directly to the high-power fields from the NMH cohorts in the same manner as described above. Training details are provided in the supplementary materials, section 1.1.

For block-level predictions we developed a model that captures both near-range and long-range interactions between fields in a memory-efficient manner, which was summarized in sections 1.2 to 1.4 and **Figures S5** to **S7**. The Sparse Convolutional Transformer (SCT) model combines self-attention and convolution using a novel element-wise self-attention (ESA) mechanism that models both spatial and channel-wise relationships (**Figure S6**). Fields are aggregated into 3x3 neighborhoods and ESA is applied to each field, followed by a within-neighborhood convolution. SCT models are composed of sequences of transformer blocks of that combine ESA and convolution operations with pooling, normalization, and multi-head self-attention that produce a multi-scale representation that combines finer details with higher-order contextual information (**Figure S7, Table S5**).

**Prediction models for cancer detection and grading**

SCT models for carcinoma detection were trained to predict block-level labels for benign / carcinoma and SCT models for grading were trained to predict block-level ISUP grade and primary and secondary Gleason pattern. The grading model contains shared layers to learn Gleason pattern morphologies followed by independent dense layers for primary and secondary pattern prediction. Appendix 1.2 to 1.4 contains additional details of the models and baselines. We used the pROC R package to evaluate statistical significance of model AUC measurements using paired and unpaired DeLong tests [28]. McNemar's test was also used to assess the significance of model sensitivities and specificities. The 95% confidence intervals for kappa statistics are also obtained for comparison.

**Dual model detection for equivocal slides**

For cancer detection in equivocal cases where IHC is typically ordered, we developed complementary SCT models: one highly sensitive model for ruling out the presence of carcinoma and one highly specific for ruling in carcinoma (**Figure S3B**). These models are combined to identify clear negative and positive blocks, and truly equivocal blocks where the models are not reliable and where IHC is required.

# Results

**Cancer detection performance**

Performance for SCT detection models on the NMH prospective validation set is shown in **Figure 2**. SCT-UNI and SCT-TSE have comparable performance with no statistically significant differences in AUC, sensitivity, or specificity. False negative errors were primarily associated with blocks containing a small cancer extent, with the large majority occurring in blocks with <10% pathologist reported cancer extent (**Figure 2B**). A t-SNE visualization of the block-level representations from SCT-TSE model shows clear clusters for benign and carcinoma blocks, and a transitional zone containing blocks with a mixture of benign and carcinoma labels (**Figure 2C**). Outliers may indicate errors in prediction or ground truth, but a pathology review of these outliers was not yet completed at the time of this publication. Disparities in detection performance were observed for self-identified Black patients who represent 7.3% of the NMH prospective dataset, although these disparities are not statistically significant when compared to self-identified White patients (**Figure 2D**). Performance was consistent across age groups with a small disparity for patients aged 50-61 (**Figure 2E**). This disparity was not statistically significant when comparing patients aged 50-61 to all other age groups. Benign / carcinoma classification AUC of the field-level classifier was 96.8% when evaluated on PANDA (see examples **Figure S8**). We compared inference time, model size, and AUC for the SCT detection model with UNI and TSE encoders in a Pareto plot (**Figure S9**). The task-specific encoder, with approximately 50 times fewer parameters, achieves 6.5 times faster inference while maintaining similar cancer detection performance (UNI: 98.7% AUC, TSE: 98.5% AUC).

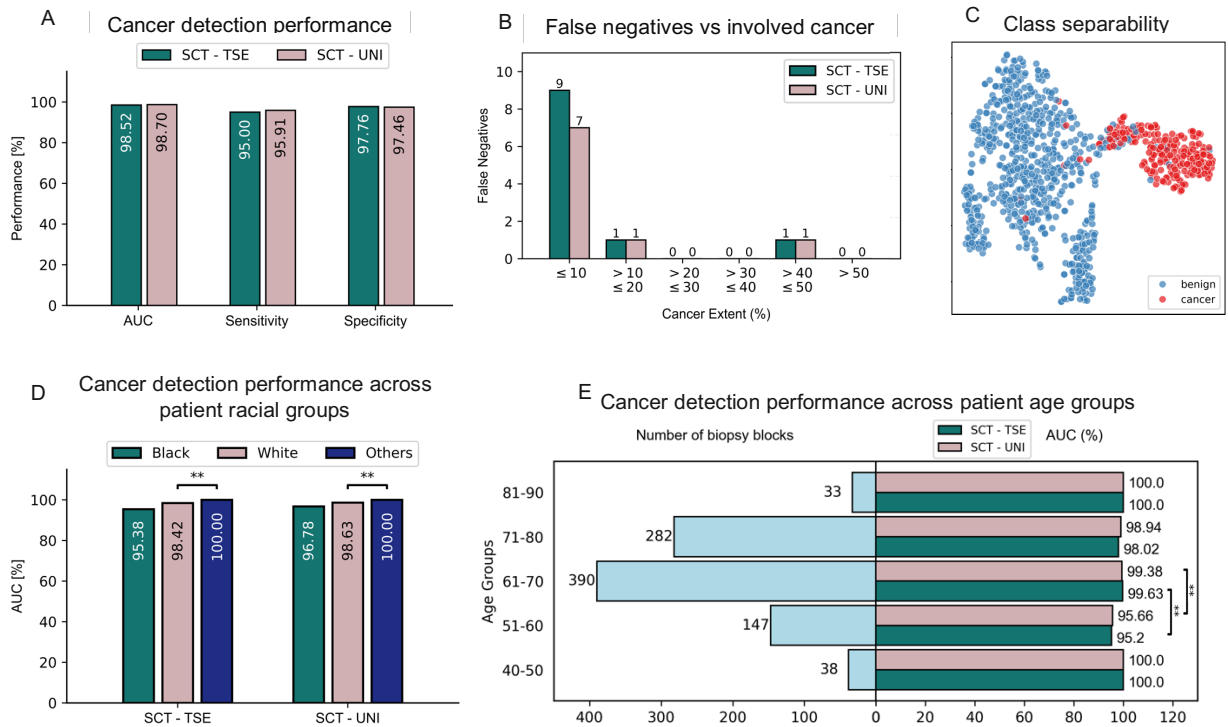

**Figure 2. Cancer detection performance.** (A) Comparison of TSE and UNI detection model performance on the prospective validation set. No statistically significant differences were observed. (B) Distribution of false negatives by cancer extent. (C) Visualization of block-level embeddings learned by SCT-TSE detection model. (D) Performance differences for self-identified Black and White patients were not statistically significant. (E) Performance by age group. The disparity in performance for patients aged 51–60 was not significant when compared to all others.

We also compared the SCT detection model to Attention-Based Multiple Instance Learning (ABMIL) and Vision Transformer (ViT) baselines (**Figure S10**) [29, 30]. For both encoders, SCT model performance is significantly better than baselines for TSE models. Visualizations of t-SNE embeddings for all models and encoders are presented in **Figure S11** along with detailed tables of errors in **Table S6**. Our cancer detection model also shows performance comparable to leading commercial tools for prostate cancer diagnosis (**Table S1**).

**Screening detection for immunohistochemical analysis**

The TSE dual-model approach screened 55% of equivocal blocks where IHC was ordered with a total error rate of 1.4%. Using a conservative threshold of 0.05, the TSE sensitive model screens 41.8% of benign blocks with a false negative rate of 1%. At a similar conservative threshold of 0.95, the TSE specific model screened 67.4% of carcinoma blocks at a false positive rate of 1.9%. In comparison, UNI achieved a higher screening rate of 82.78% but with a higher total error rate of 3.8%. The UNI sensitive model screened 60.1% of benign blocks with a false negative rate of 3.8%, while the UNI specific model screened 69.6% of carcinoma blocks with a false positive rate of 4%. Despite higher screening rates, UNI models have uniformly higher error rates than TSE models at all thresholds. The trade-offs between confidence thresholds and the screening performance and error rates of the dual-model approach with TSE encoder are illustrated in **Figure 3**. Relaxing these thresholds screens a larger proportion of cases but with increased error rates. **Figure 4A-B** visualize classification boundaries of the dual models with TSE at various thresholds.

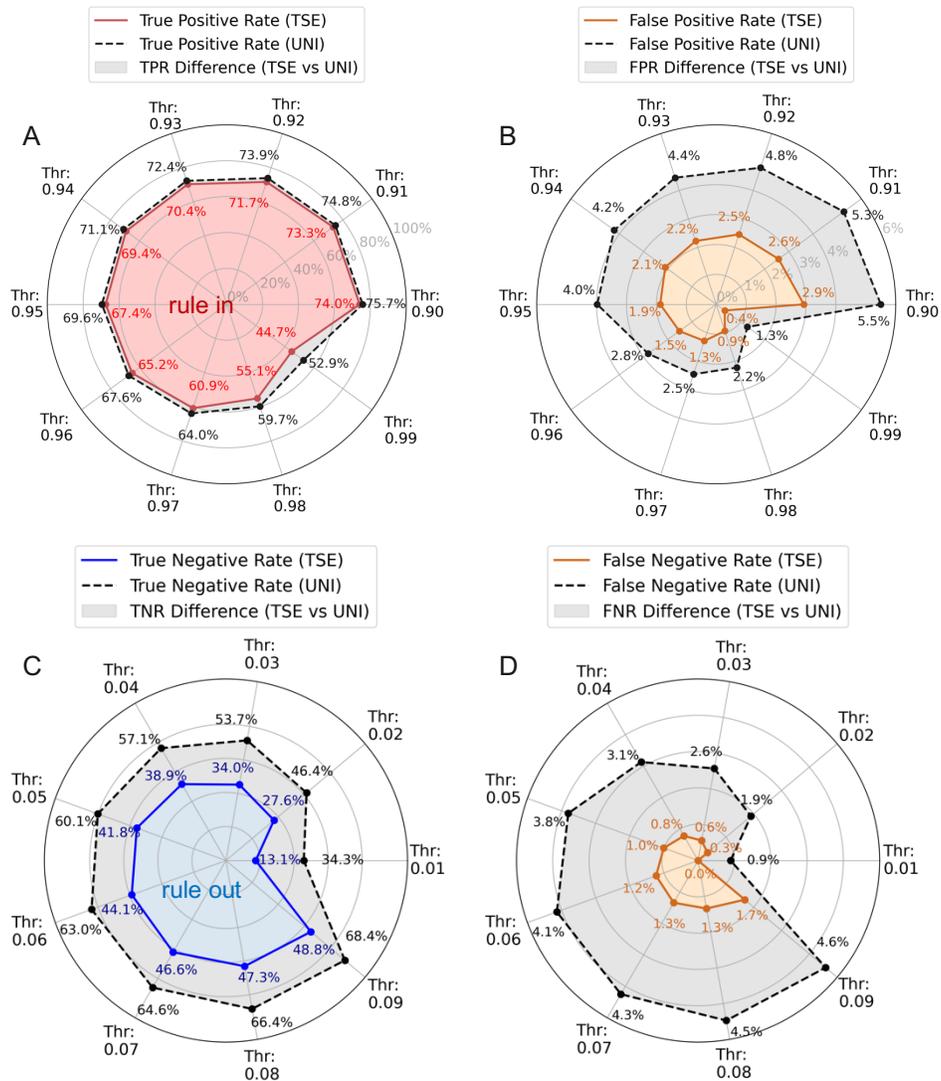

**Figure 3. Immunohistochemical screening performance for TSE and UNI.** Trade-offs between model decision thresholds and true positive rates (TPRs) (A), false positive rates (FPRs) (B), true negative rates (TNRs) (C), and false negative rates (FNRs) (D). Fractions of screened carcinoma blocks ("rule in") as a function of thresholds are depicted in A with shaded red, while screened benign blocks ("rule out") are shown as shaded blue in C. Errors (FNR and FPR) depicted in orange can be reduced using stricter thresholds at the cost of additional IHC ordering. For instance, at a threshold of 0.99, the TSE models screen 44.7% of carcinoma and 13.1% of benign blocks at FPR and FNR of 0.4% and 0%, respectively, while the UNI models screen 52.9% of carcinoma and 34.3% of benign blocks at FPR and FNR of 1.3% and 0.9%, respectively.

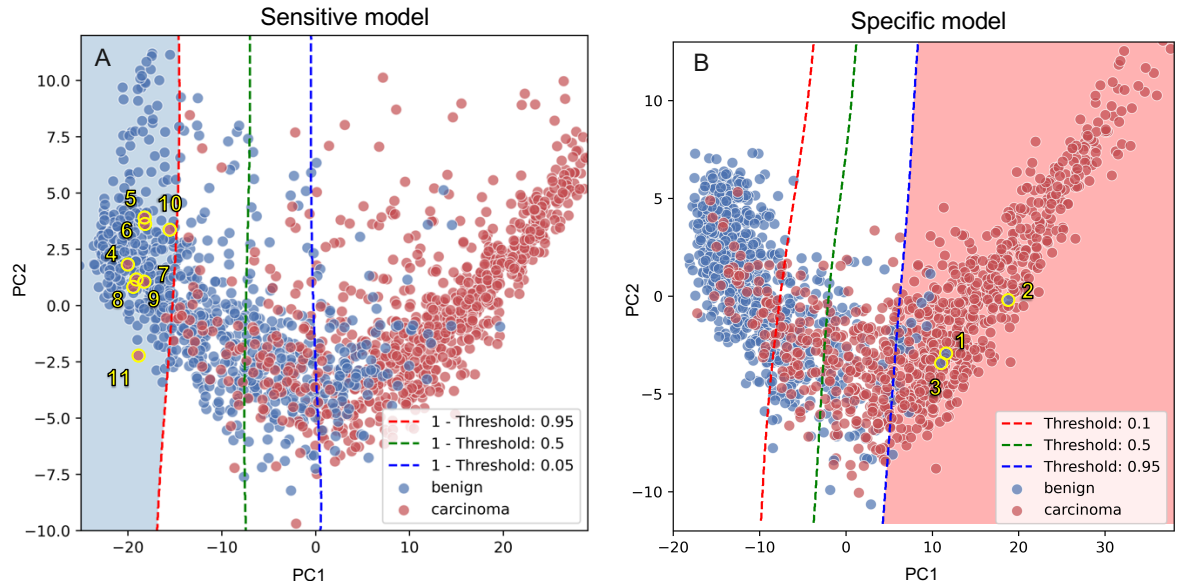

**Figure 4. Dual model classification with TSE visualized by principal component analysis**. Each point represents the low dimensional feature embedding learned for a single tissue block. (A) Classification boundaries for the sensitive model at thresholds of 0.95, 0.5, and 0.05. The shaded blue area corresponding to a strict threshold of 0.05 (1-0.95) that rules out 41.8% of benign blocks with a false negative rate of less than 1%. (B) At the threshold of 0.95, the specific model rules out 67.8% of carcinoma blocks (red shaded area) with a false positive rate of 1.9%. The visualizations are low-dimensional representations for illustration purposes and do not necessarily represent separability in the unobservable higher-dimensional space.

Using this plot, we selected 11 misclassified outlier tissue blocks from as many blocks (highlighted with yellow circles) for consensus review by a panel of five genitourinary fellowship-trained pathologists (see section 1.5 and **Table S7** in the supplementary document). The initial pathologic diagnosis for blocks 1 and 3 identified atypical small acinar proliferation (ASAP) and prostatic intraepithelial neoplasia with adjacent small atypical glands (PINATYP) but no GG1 cancers. Consensus review of these blocks identified small volumes of GG1 cancer, consistent with the model predictions. Block 2 was also found to contain a small GG1 cancer, but one other block from that patient also contained a GG1 cancer, and so this would not change the diagnosis. For blocks 5-11 where the model predicted benign, the consensus review determined that the model was incorrect. Consensus review found that these blocks include difficult patterns including atrophic cancer, ductal carcinoma, cribriform growth pattern, and tissue processing artifacts (e.g. dehydration). Block 4 exhibited atrophic features with high-grade prostatic intraepithelial neoplasia (HGPIN) and ASAP, and although a few cancerous glands were identified, the review concluded that this was insufficient for conclusive diagnosis. With the consensus diagnoses as the reference, the specific model demonstrates enhanced screening capabilities, identifying 67.6% of carcinoma blocks with a minimal FPR of 1.5% using TSE. **Figure S12** also illustrates tissue sections from the reviewed blocks, highlighting the regions of interest identified by the consensus panel.

**Cancer grading performance**

Performance of the SCT grading model is presented in **Figure 5**. SCT-TSE and SCT-UNI have similar grading performance with significant differences observed in prediction of secondary

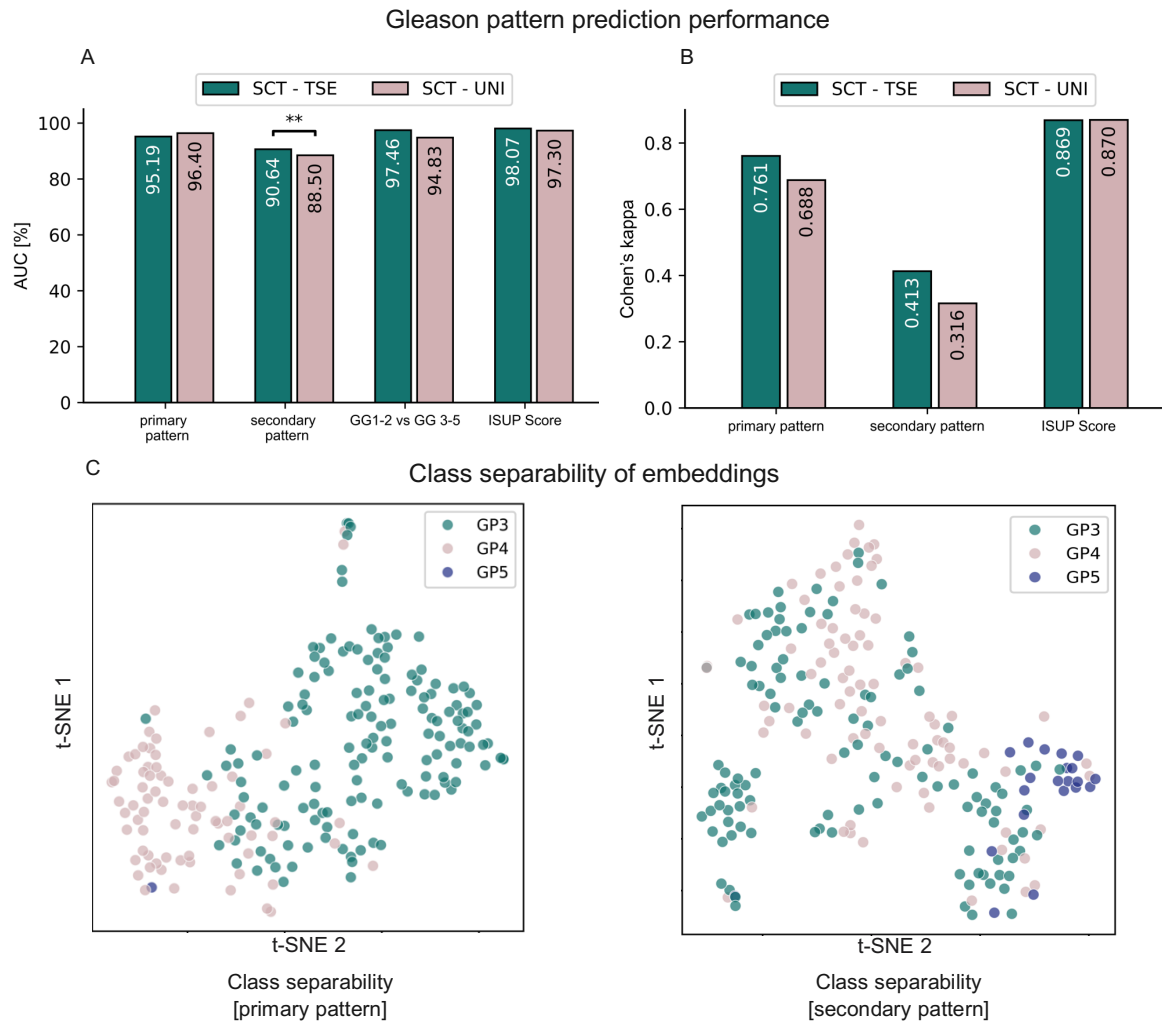

**Figure 5. Gleason grading model performance**. (A) Performance of block-level predictions for primary and secondary Gleason patterns, ISUP scores, and discrimination between Gleason GG 1-2 versus GG 3-5 using the SCT grading model with task-specific and universal features. (B) Kappa statistics for agreement between predicted and actual primary and secondary Gleason patterns, and ISUP scores. (C) Visualization of class separability for latent features learned by SCT grading model with the task-specific encoder for primary and secondary block-level patterns. Significance levels are denoted by asterisks with ** for $p < 0.01$.

Gleason pattern (**Figure 5A**). The AUC for ISUP grade prediction ranges from 97.3% (SCT-UNI) to 98.1% (SCT-TSE) with corresponding kappa values of 0.869 (95% CI: 0.836 to 0.901) and 0.870 (95% CI: 0.837 to 0.902) respectively. Prediction for the clinically important threshold of GG1-2 versus GG3-5 has AUCs of 97.5% (SCT-TSE) and 94.8 (SCT-UNI). Larger differences between TSE and UNI were observed in kappa values for direct prediction of primary and secondary Gleason patterns (**Figure 5B**). SCT-TSE kappa values were generally higher than those of SCT-UNI with 0.761 (95% CI: 0.675 to 0.847) versus 0.688 (95% CI: 0.592 to 0.784) for primary pattern and 0.413 (95% CI: 0.292 to 0.534) versus 0.317 (95% CI: 0.191 to 0.442) for secondary pattern. A t-SNE visualization of the SCT-TSE grading model illustrates separation of blocks by primary Gleason patterns. In contrast, although some clusters for secondary patterns

Page **9** of **13**

were present, they were less distinct and compact, suggesting a higher degree of subjectivity in secondary pattern prediction (**Figure 5C**).

A grading performance comparison between SCT and baselines is shown in **Figure S13**. SCT models perform better than baselines in grading tasks with the exception of SCT-UNI for secondary pattern prediction. The SCT-TSE and SCT-UNI grading models attained AUCs of 95.2% and 96.4% for primary patterns and 90.6% and 88.49% for secondary patterns, respectively, outperforming ViT (94.56%, 89.93%) and ABMIL (93.54%, 88.54%). Additionally, SCT-TSE and SCT-UNI grading models achieved higher quadratic Cohen's kappa scores of (0.761, 0.688) for primary Gleason pattern, showcasing better performance compared to ViT (0.676, 0.532) and ABMIL (0.695, 0.643) models. For the secondary Gleason pattern, SCT grading models also showed better performance, achieving kappa scores (0.413, 0.317) with task-specific and UNI encoders. As presented in **Table S2**, our model also exhibits grading performance equivalent to that of prominent commercial tools for prostate cancer grading.

## Discussion

Pathologists face increasing challenges delivering timely diagnosis for prostate biopsies. AI offers a possible solution to reduce turnaround times, improve quality, and reduce costs by improving pathologist efficiency and reducing use of immunohistochemical staining. Our study investigated the development of an institutional AI system for prostate cancer detection, grading, and optimization of IHC ordering. This work presents novel modeling approaches, a comparison of large foundation encoder with a more efficient task-specific alternative, and real-world validation. We demonstrate that this system could potentially reduce use of immunohistochemical staining and has comparable performance in cancer detection and grading with published validations of commercial systems. Our study also incorporated a consensus review of cases where IHC was ordered that identified cancers not initially diagnosed by pathologists.

By combining complementary sensitive and specific models we show that our algorithms can correctly classify 55% overall IHC ordering needs that occurred at our institution with an error rate of 1.4% for equivocal cases where immunohistochemistry was ordered. In practice, IHC will remain the gold standard for positive cancer diagnosis, and so in reality future benefits to workflow and efficiency would come from the sensitive model ruling out negative blocks (41.8% of benign blocks, 1% error rate in this case). Any reduction, however, can reduce costs and unnecessary anxiety for patients who undergo biopsy.

The consensus review identified three blocks from as many cases where the model prediction of cancer was correct and discordant with the initial pathologic diagnosis at the block or patient level. For two of these cases, the initial diagnosis of ASAP indicates a suspicion of cancer but too small to be conclusive, and so a follow-up biopsy would have been recommended for these patients regardless. Detection of a low-grade cancer would not have resulted in a change in clinical management for these patients. The biopsy of the third case contained an additional positive core of the same grade, and so the patient's diagnosis and risk group would not have changed. Although this patient would likely be managed by active surveillance, in general it is important to detect all positive cores within a biopsy, since some patients may be eligible for focal therapy. The review also confirmed many negatives by the model that contained challenging patterns like benign mimickers and tissue processing artifacts.

Our comparison of a task-specific and foundation encoders shows indistinguishable performance in cancer detection, with the foundation model achieving sensitivity and specificity of 95.91% and 97.46%, and TSE achieving 95% and 97.76%, respectively, but some marginal advantage for TSE in grading. Foundation models have gained significant popularity due to their versatility and strong performance in many tasks, particularly where training data is lacking. Our analysis compares foundation models to smaller and more efficient task-specific alternatives in prostate applications where data are abundant. We also make this comparison in the context of IHC screening, an application that differs from traditional classification benchmarks. Screening of IHC cases in prostate requires analyzing the most difficult slides where the pathologist requires assistance, and places strict constraints on error rates for accurate screening. These patterns may not be as easily captured by foundation models that emphasize broader distinctions. The narrow focus of task-specific encoders improves their performance in screening, aided by the availability of abundant prostate data for training. The small size of the TSE also significantly reduces the time and costs of model inference in deployment, with approximately 50 times fewer parameters and 6.5 times faster inference, while maintaining comparable cancer detection performance (UNI foundation model: 98.7% AUC, TSE: 98.5% AUC) TSE development, however, was non-trivial requiring supervised training using a large public dataset containing carcinoma annotations.

We developed SCT to address the specific challenges of cancer detection and grading in prostate biopsies. Detection of small cancer foci and Gleason pattern grading should benefit from modelling spatial-contextual relationships between high-power fields. Gleason patterns are more apparent at lower power, and the presence of adjacent fields containing suspicious patterns can raise the model's confidence in predicting cancer. While ABMIL lacks the capacity to model interactions, ViT models interactions between all fields. In contrast, SCT uses an efficient neighborhood processing approach that progressively expands the scope of modeled interactions, reducing computation and trainable parameters compared to ViT. This results in superior performance for SCT in almost all comparisons we examined.

Our study has important limitations. Our goal was to develop models for internal use, and to understand their performance if deployed within our hospital, and so we did not validate on external datasets. However, our validation data were collected prospectively without filtering or enrichment of cases based on grade or cancer burden and represent an unbiased sample of our patient population. We expect that similar results could be obtained by other institutions developing and validating models internally, however, performance may vary with the degree of preanalytical variability or differences in patient populations. An analysis of racial disparities in cancer grading performance was not possible due to the size of our validation dataset. Our analysis of foundation model encoders did not include models beyond UNI or perform additional fine tuning. Future research will study this system prospectively in clinical deployment to analyze potential impacts on quality, efficiency, and costs.

## Acknowledgements

This work was funded by the National Institutes of Health National Library of Medicine award R01LM013523 and the Northwestern Polsky Urological Cancer Institute. We would like to acknowledge the Information Technology team at Northwestern University and the Electronic Data Warehouse team at Northwestern Medicine for their support. The Northwestern Electronic Data Warehouse is supported by NUCATS grant UL1TR001422 from the National Institutes of Health's National Center for Advancing Translational Sciences. HDP is supported by a Prostate

Cancer Foundation Young Investigator Award and a Developmental Research Program grant from the SPORE in Prostate Cancer (P50 CA 180995).## Contributors

RN was responsible for the conceptualisation, methodology development, formal analysis, visualisation, validation, and drafting of the original manuscript, as well as for its editing. RZ, MSa, and MS played key roles in data acquisition and curation, with CN providing additional support in data curation. MSa also contributed to writing the original introduction. RA, NH, MH, EVL, JAG, HDP, and MN provided insights through manuscript review and editing. EMS led the project administration. FP was involved in visualisation. The validation and investigation efforts were led by pathologists BI, CF, VM, BGN, and XJY, all of whom also contributed to manuscript review and editing. AER made contributions by providing resources, supervising the project, managing project administration, securing funding, conceptualisation, and participating in manuscript review and editing. LADC supported the conceptualisation, supervision, project administration, resource provision, and manuscript review and editing.

## Declaration of interests

HDP received a Young Investigator Award for AI in prostate cancer imaging from the Prostate Cancer Foundation, unrelated to the present study. Additionally, the author participated in a one-time Advisory Board meeting in 2024 for Cleveland Diagnostics, a company that developed a prostate cancer diagnostic test, with no ongoing relationship or relevance to the current research. The other authors declare no conflicts of interest.

## References

1. Sekhoacha, M., et al., *Prostate Cancer Review: Genetics, Diagnosis, Treatment Options, and Alternative Approaches*. Molecules, 2022. **27**(17).
2. Quon, H., A. Loblaw, and R. Nam, *Dramatic increase in prostate cancer cases by 2021*. BJU Int, 2011. **108**(11): p. 1734-8.
3. James, N.D., et al., *The Lancet Commission on prostate cancer: planning for the surge in cases*. Lancet, 2024. **403**(10437): p. 1683-1722.
4. Rozario, S.Y., et al., *Responding to the healthcare workforce shortage: A scoping review exploring anatomical pathologists' professional identities over time*. Anat Sci Educ, 2024. **17**(2): p. 351-365.
5. Raciti, P., et al., *Novel artificial intelligence system increases the detection of prostate cancer in whole slide images of core needle biopsies*. Mod Pathol, 2020. **33**(10): p. 2058-2066.
6. Pantanowitz, L., et al., *An artificial intelligence algorithm for prostate cancer diagnosis in whole slide images of core needle biopsies: a blinded clinical validation and deployment study*. Lancet Digit Health, 2020. **2**(8): p. e407-e416.
7. Sooriakumaran, P., et al., *Gleason scoring varies among pathologists and this affects clinical risk in patients with prostate cancer*. Clin Oncol (R Coll Radiol), 2005. **17**(8): p. 655-8.
8. Al-Maghrabi, J.A., N.A. Bakshi, and H.M. Farsi, *Gleason grading of prostate cancer in needle core biopsies: a comparison of general and urologic pathologists*. Ann Saudi Med, 2013. **33**(1): p. 40-4.Page **12** of 13

# 1 Supplementary methods

## 1.1 Field-wise carcinoma delineation and Gleason mapping.

To develop a task-specific encoder for prostate cancer, we first train a high power field classifier of benign/carcinoma using PANDA data which comprises of 10616 biopsy slides obtained from Radboud University Medical Center and Karolinska Institute (**Table S**4). The PANDA images were tiled into 256x256 pixel high-power fields at 20X objective magnification. The CNN encoder, parameterized by $\alpha$, is learned to extract high-level d-dimensional features $f^{(i)} = \emptyset_\alpha(t^{(i)})$ from input tiles $t^{(i)}$, which enables the classification layers $\varphi_\beta(f^{(i)})$ parameterized by $\beta$ to identify carcinoma across slides. As the NU dataset lacks detailed region-level annotations, we utilize tiles from the PANDA dataset for training. The encoder and classification layers are trained based on the training samples $D_P = \{s^{(i)} = (t^{(i)}, y^{(i)}) \mid i = 1, 2, 3, ..., N\}$, where $s^{(i)}$ is the $i^{th}$ training sample, and $y^{(i)}$ is the tile-level label. To make the model better at ignoring benign mimickers in our dataset, we proceed with a fine-tuning phase following the initial training of the classification layers and encoder. First, we apply the model to purely benign slides in NU dataset to identify misclassified benign tiles. Given that all the tiles from benign slides are expected to be entirely benign, we employ the misclassified benign tiles for targeted refinement in a process known as hard mining. To retain the general knowledge learned in the pre-trained model, we only make the last three convolution layers of the encoder trainable, which helps in tuning the model against false detections while preserving generalizability.

The NMH dataset lacks image annotations, and so we evaluated the quality of encodings by visualizing the benign / carcinoma classifications on NMH whole-slide images. We also trained a similar high-power field classifier for Gleason patterns but did not use this for purposes other than visualization.

## 1.2 Tile indexing layer

Following the feature extraction step, where features from individual tiles across all slides within a block are extracted and subsequently concatenated into a 2D sparse matrix, given that SCT integrates convolutional operations and local attention mechanisms, the model needs to know the spatial arrangement of tiles within each slide of a block. Therefore, we integrate three inputs into the model: features extracted from tiles $F \in \mathbb{R}^{N \times D}$, their corresponding coordinates $C = \{(x_i, y_i) \mid i: 1, 2, ... N\} \in \mathbb{R}^{N \times 2}$, and unique indices $I = \{s_i \mid i: 1, 2, ... N\} \in \mathbb{R}^{N \times 1}$ representing the originating slide for each tile containing values starting at 1, with 1 representing the first slide, 2 for the second slide, and so on. The tile indexing layer plays a crucial role in our model ensuring that the coordinates of tiles across all slides within a tissue block are uniquely encoded, addressing the potential ambiguity that arises when multiple tiles from different slides share the same coordinates. **Figure S5** represents the pseudo-code for tile indexing. First, we organize the tiles based on which slide they come from. Given tile coordinates $C$ and tile indices $I$, we group tiles by slide using $C_j = \{(x_i, y_i) \mid s_i = j\}$ where $C_j$ represents the set of the tile coordinates belonging to $j^{th}$ slide. For each slide $j$, we find the maximum coordinates $(X_j, Y_j)$ in both the x and y directions, where $X_j = \max\{x_i \mid (x_i, y_i) \in C_j\}$, and $Y_j = \max\{y_i \mid (x_i, y_i) \in C_j\}$. This means we find the farthest right and farthest down that any tile goes on that slide. After finding the maximum coordinates for each slide, we adjust the coordinates of subsequent slides based on the maximum coordinates of the previous slide as $C'_j = \{(x_i, y_i) \mid s_i = j\} + (X_{j-1}, Y_{j-1})$, where $C'_j$ is the adjusted coordinates for $j^{th}$ slide. By doing this, we make sure that no two tiles have the same coordinates across all slides in a tissue block.

## 1.3 Sparse Convolutional Transformer (SCT)

We introduce the Sparse Convolutional Transformer (SCT), an innovative memory-efficient model tailored for the processing of gigapixel digital pathology slides. Unlike conventional vision transformer models, SCT seamlessly integrates localized details and global patterns through its adept utilization of local and global self-attention modules. This section provides a comprehensive overview of the SCT architecture and its distinctive features.

The SCT block takes adjusted tile coordinates $C' \in \mathbb{R}^{N \times 2}$ and extracted features tokens $F \in \mathbb{R}^{N \times D}$ as inputs. Initially, the tiles level features tokens are projected to a lower dimension $Z$ using a linear fully connected layer parameterized by $W \in \mathbb{R}^{Z \times D}$ and then normalized by a layer normalization to prevent unbiased training due to unexpected higher value features. The resulting features and corresponding coordinates are then pass through a Spatially Sparse Convolutional Self-Attention (SSC-SA) block, which combines the efficiency of convolutional operations with the relationship-modeling capability of self-attention and effectively captures localized spatial and channel-wise dependencies within the token receptive fields. A receptive field defines a 2D region of neighboring tokens determined by a kernel with a size of $k \times k$, and a stride ratio of $s$, which defines how the model looks at neighboring tokens when performing convolutional and self-attention operations. A skip connection is applied from the input to the output of the SSC-SA block to help the model learn more complex representations while maintaining efficient gradient flow. We introduce a Spatially Sparse Pooling (SSP) block is then employed to reduce the number of tokens. It aggregates information from neighboring tokens with receptive fields, which helps in the extraction of key information while reducing computational complexity. After normalizing the resulting feature tokens, they are then passed through a global multi-head self-attention layer After normalizing the resulting feature tokens, the feature query, key, and value are then passed through a global multi-head self-attention layer, enabling interaction between feature tokens across all slides of a block. Finally, after the features are normalized, they proceed through a Multi-Layer Perceptron (MLP) layer, consisting of two fully connected layers. These layers are characterized by $W_1 \in \mathbb{R}^{Z \times Z}$ and $W_2 \in \mathbb{R}^{Z \times Z}$ and utilize the Gaussian Error Linear Unit (GELU) activation function. to facilitate the flow of information and gradients throughout the network, we also added two skip connections between the inputs and outputs of both MLP and the global multi-head self-attention layers.

### 1.3.1 Spatially Sparse Convolutional Self-Attention (SSC-SA)

The sparse convolutional operation is applied to capture local spatial dependencies between neighboring tokens within receptive fields, which is summarized in **Figure S6A** Given a set of feature tokens $F \in \mathbb{R}^{N \times D}$ and corresponding coordinates $C' = \{(x_i, y_i) \mid i: 1, 2, \ldots N\} \in \mathbb{R}^{N \times 2}$, we initially generate receptive fields for each token by considering its neighboring tokens within the specified kernel with size of $k \times k$. This generates a 2D region of neighboring tokens centered around each token of interest. Given a feature token $f^{(i)} \in \mathbb{R}^{1 \times Z}$ and corresponding coordinates $(x_i, y_i)$, the receptive field $R_i$ is defined as the set of tokens located within neighboring $k \times k$ kernel.

$$R_i \in \mathbb{R}^{k^2 \times Z} = \left\{ f^{(j)} \,\middle|\, |x_j - x_i| < \left\lfloor \frac{k}{2} \right\rfloor, |y_j - y_i| < \left\lfloor \frac{k}{2} \right\rfloor, j = 1, 2, 3, \ldots N \right\}$$

We introduce an Element-wise Self-Attention (ESA) mechanism that is applied to the receptive fields to further refine the representations and capture local spatial and feature dependencies across tokens within receptive fields, which is summarized in **Figure S6B** Given receptive field $R_i \in \mathbb{R}^{k^2 \times Z}$, we initially leverage a trainable fully connected layer with a weight $W_r \in \mathbb{R}^{Z \times C}$ to reduce the dimensionality of $R_i$, where $C$ is the embedded feature dimension. Considering the resulting receptive field as query $Q_i \in \mathbb{R}^{k^2 \times C}$, key $K_i \in \mathbb{R}^{k^2 \times C}$, and value $V_i \in \mathbb{R}^{k^2 \times C}$ we calculate the attended spatial representations $Y_i \in \mathbb{R}^{k^2 \times C}$ using a softmax normalization as below:

$$A_i \in \mathbb{R}^{k^2 \times k^2} = \text{softmax}\left(\frac{Q_i \cdot K_i}{\sqrt{C}}\right)$$

$$Y_i = A_i \cdot V_i$$

Where $A_i$ represents the attention weights between local feature tokens within a receptive field $R_i$. The next step is to apply attention in the channel direction. To achieve this, we first transpose the input receptive field matrix to ensure that the attention mechanism operates along the channel dimension. Transposing the input receptive field matrix reorganizes the features such that the channels become the new rows and the spatial dimensions become the new columns. This transformation allows the attention mechanism to compute relationships between features across different channels. Considering the transposed receptive field as query $Q'_i \in \mathbb{R}^{C \times k^2}$, key $K'_i \in \mathbb{R}^{C \times k^2}$, and value $V'_i \in \mathbb{R}^{C \times k^2}$ we compute the attended feature representation $Y'_i \in \mathbb{R}^{k^2 \times C}$ as below.

$$A'_i \in \mathbb{R}^{C \times C} = \text{softmax}\left(\frac{Q'_i \cdot K'_i}{\sqrt{k^2}}\right)$$

$$Y'_i = A'_i \cdot V'_i$$

The resulting $Y'_i$ is then transposed and combined with the attended spatial representations $Y_i$, and subsequently, a trainable fully connected layer with a weight $W_o \in \mathbb{R}^{C \times Z}$ is employed to return the dimension back to the input dimension as $R'_i \in \mathbb{R}^{k^2 \times Z} = (Y'_i + Y_i) W_o$. We also added a skip connection between the input and output of the ESA block to ensure generalizability.

Once the attended receptive field is obtained, the next step involves the application of convolution to each receptive field using a convolution weight $W_c \in \mathbb{R}^{D \times k^2 \times Z}$.

$$f'^{(i)} = \text{conv}(R'_i, W_c) = \sum_{\substack{j=1 \\ f^{(j)} \in R'_i}}^{k^2} f^{(j)} \cdot W_c$$

where $D$ is the output dimension (the number of filters in the convolution) and $f'^{(i)}$ represents the resulting token obtained from SSC-SA block for the $i^{th}$ receptive field centered around $(x_i, y_i)$.

### 1.3.2 Spatially Sparse Pooling (SSP)

The SSP layer is a key component of the SCT block, as it gathers information from neighboring tokens through receptive fields and enables the extraction of important details while minimizing computational complexity and providing multi-scale representations. In the SSP layer, the pooling operation aggregates information from neighboring tokens within receptive fields. These receptive fields are constructed based on the specified pool size and stride. We define a receptive field as neighboring tokens centered around each token within a specified spatial range. Given the pooling size of $p$ and stride of $s$, a receptive field $R_i \in \mathbb{R}^{k^2 \times Z}$ is calculated as below.

$$R_i \in \mathbb{R}^{k^2 \times Z} = \left\{ f^{(j)} \middle| |x_j - x_i| < \left\lfloor \frac{p}{2} \right\rfloor, |y_j - y_i| < \left\lfloor \frac{p}{2} \right\rfloor, j = 1, 2, 3, \ldots N, i \in S \right\}$$

$$S = \{i \mid x_i \in X \text{ with stride } s, y_i \in Y \text{ with stride } s\}$$

Where $X$ represents the set of possible $x$-coordinates and $Y$ represents the set of possible $y$-coordinates, and S contains the indices $i$ where both the $x_i$ and $y_i$ coordinates lie within the 2D space defined by $X$ and $Y$, with a stride $s$.

Once the receptive fields are formed, we perform pooling to aggregate information from the tokens within each receptive field. The pooling operation typically aggregates tokens through a maximum or average operation across the tokens in a receptive field. For instance, if we consider max pooling, the $m^{th}$ element of pooled feature $f_p^{(i)}$ for token $i$ located at $(x_i, y_i)$ can be calculated as below.

$$f_p^{(i)}[m] = \max_{r=1}^{k^2} f_p^{(r)}[m]$$

### 1.4 Network architecture

We build models called tiny-SCT, small-SCT, and large-SCT with different parameters by modifying the number of Transformer blocks and the embedding feature dimensions, as detailed in **Table S5**.

**1.5 Consensus review of IHC cases**

We reviewed 11 misclassified biopsy blocks with a panel of five genitourinary fellowship-trained pathologists to investigate the underlying reasons for their distinct presentation in feature space (**Figure 4**). The review was conducted as a conference where the pathologists attended in person and reviewed physical slides. At the outset, they were not informed about the initial diagnosis, but after reaching a consensus, they were provided with the initial diagnosis and the model's results. The pathologists were also open to discussion, expressing agreement or disagreement with the findings as they reviewed the slides.

## 2 Supplementary figures and tables

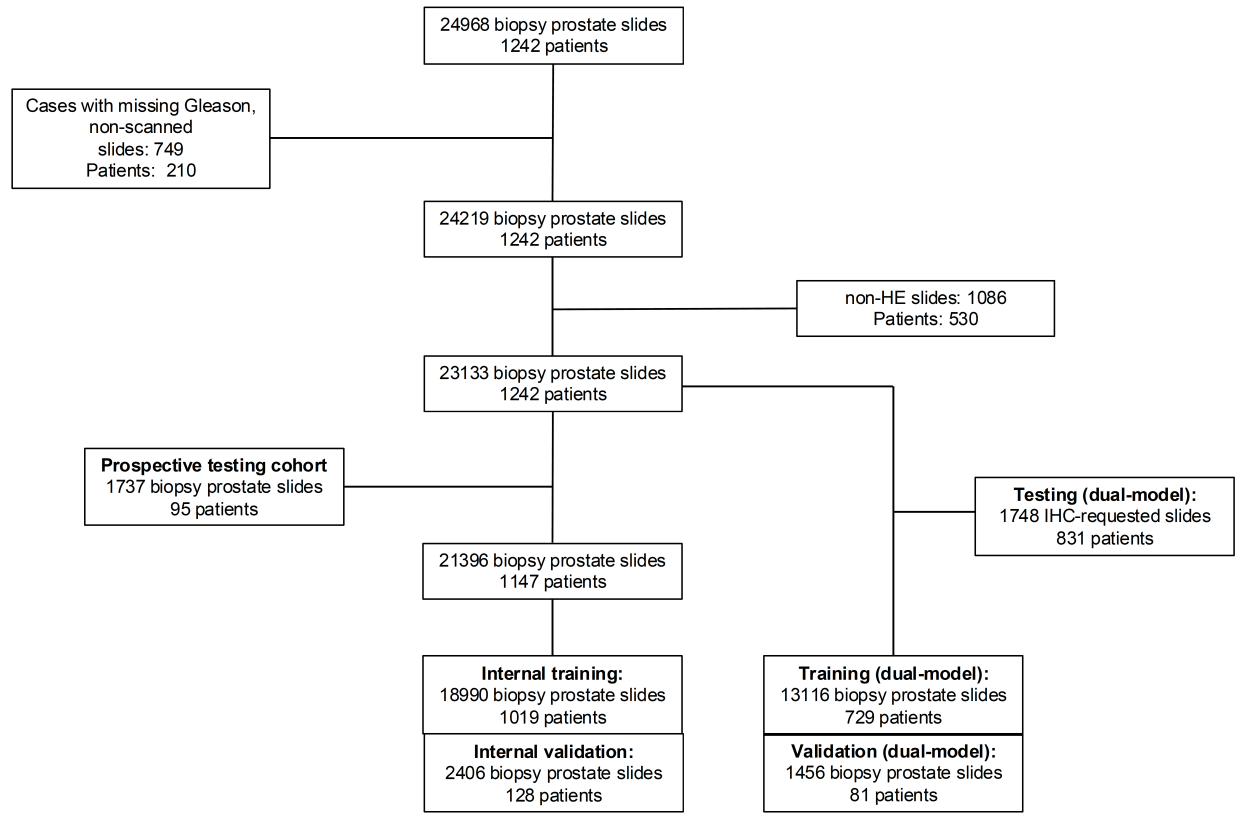

**Figure S1. Patient selection flowchart.** The dataset used to develop our cancer detection and grading models involved 24,968 prostate biopsy slides from 1,242 patients collected at Northwestern Memorial Hospital (NMH) between 2021-2023. After excluding 749 slides due to missing Gleason scores or non-scanned status and removing 1,086 non-H&E stained slides, 23,133 H&E-stained slides from 1,242 patients remained. From this, 1,737 slides collected in March and April of 2023 were designated for the prospective evaluation. The final dataset was randomized and stratified by patient into training (18,990 slides), and validation (2,406 slides) groups. For the dual-model development, a subset of the entire prostate cancer cohort was used, with slides containing IHC staining excluded, resulting in 14,572 slides. These were divided into training (13,116 slides) and validation (1,456 slides) sets with a 90:10 ratio. The IHC-requested slides, totaling 1,748, were set aside for testing.

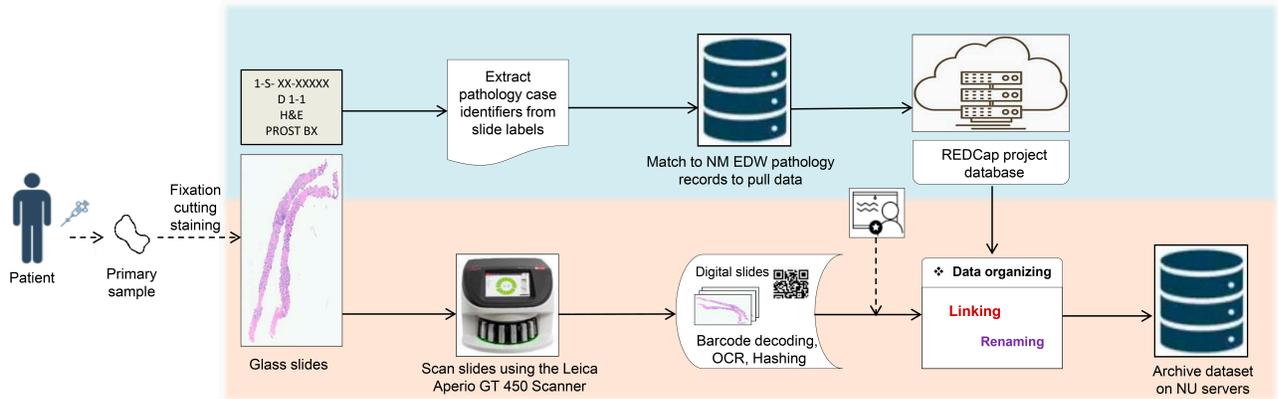

**Figure S2. Our data scanning, archiving and linking pipeline.**

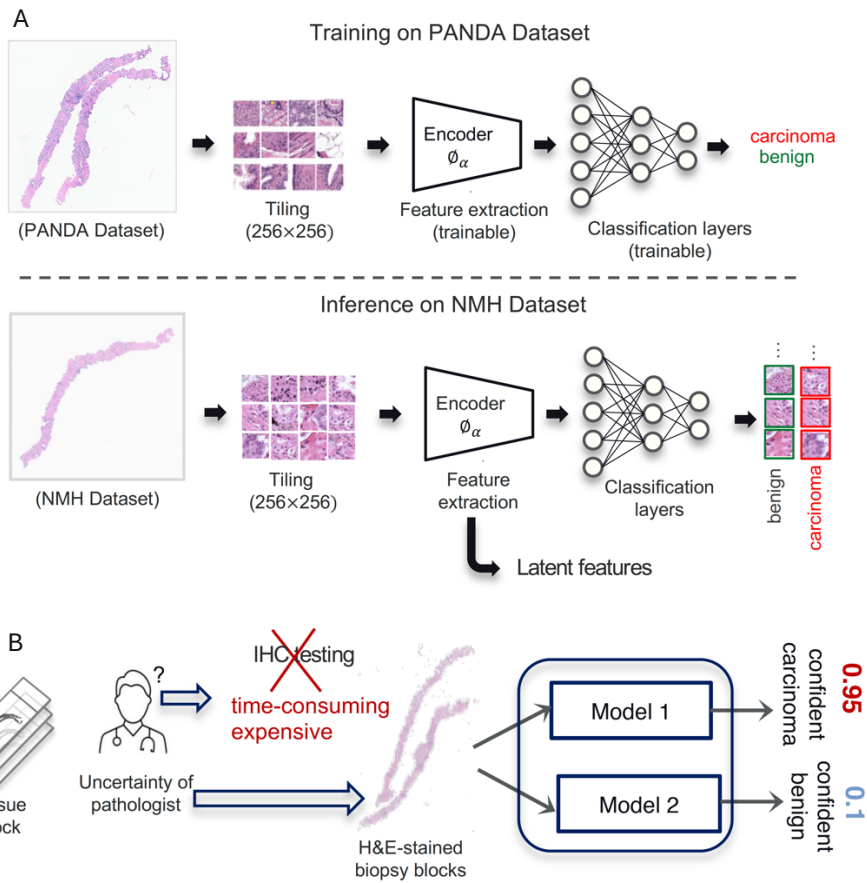

**Figure S3. Modeling approaches.** (A) A task-specific encoder based on the EfficientNetB0 network was trained to classify high-power fields using the PANDA dataset. Features from this encoder were used as inputs for block-level cancer detection and grading prediction models. (B) We use a dual modeling approach that combines highly specific and highly sensitive models to identify cases where IHC is not required with the goal of improving diagnostic efficiency and cost.

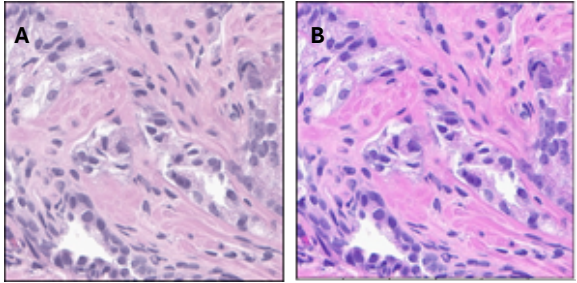

**Figure S4. ICC color profile correction.** (A) Original image, (B) ICC corrected image.

**Algorithm: Tile indexing**

| | |
|---|---|
| **Inputs** | tile coordinates: $C = \{(x_i, y_i) \mid i: 1, 2, \ldots N\} \in \mathbb{R}^{N \times 2}$ |
| | tile indices: $I = \{s_i \mid i: 1, 2, \ldots N\} \in \mathbb{R}^{N \times 1}$ |
| | features: $F \in \mathbb{R}^{N \times D}$ |
| | $N_s$: number of slides within the block |
| **Output** | adjusted unique coordinates $C' \in \mathbb{R}^{N \times 2}$ |
| | features are outputted unchangeably. |
| **Procedure** | 1. Initialize slide index $j = 0$ |
| | 2. Initialize maximum coordinate lists in $x$ and y direction: $X = Y = []$ |
| | 3. **repeat** |
| | 4.     $j \leftarrow j + 1$ |
| | 5.     group tiles by slide: $C_j = \{(x_i, y_i) \mid s_i = j\}$ |
| | 6.     $X[j] = \max\{x_i \mid (x_i, y_i) \in C_j\}$ |
| | 7.     $Y[j] = \max\{y_i \mid (x_i, y_i) \in C_j\}$ |
| | 8.     **If** $j = 1$ **then** |
| | 9.         $C_j = C_j$ |
| | 10.    **else** |
| | 11.        $C_j = C_j + (X[j-1], Y[j-1])$ |
| | 12.   **end If** |
| | 13. **until** $j = N_s$ |

**Figure S5. Pseudo code of the proposed tile indexing method.**

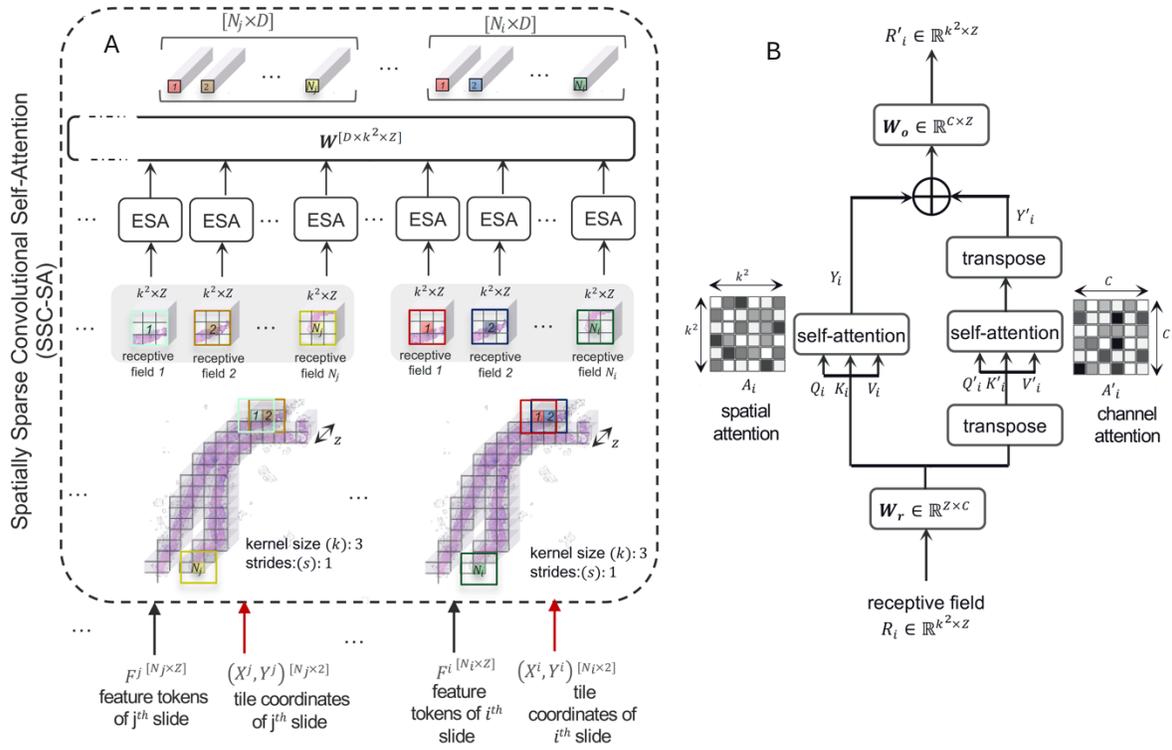

**Figure S6. Spatially sparse convolutional self-attention operation.** (A) SSC-SA Block. (B) Element-wise Self-Attention (ESA).

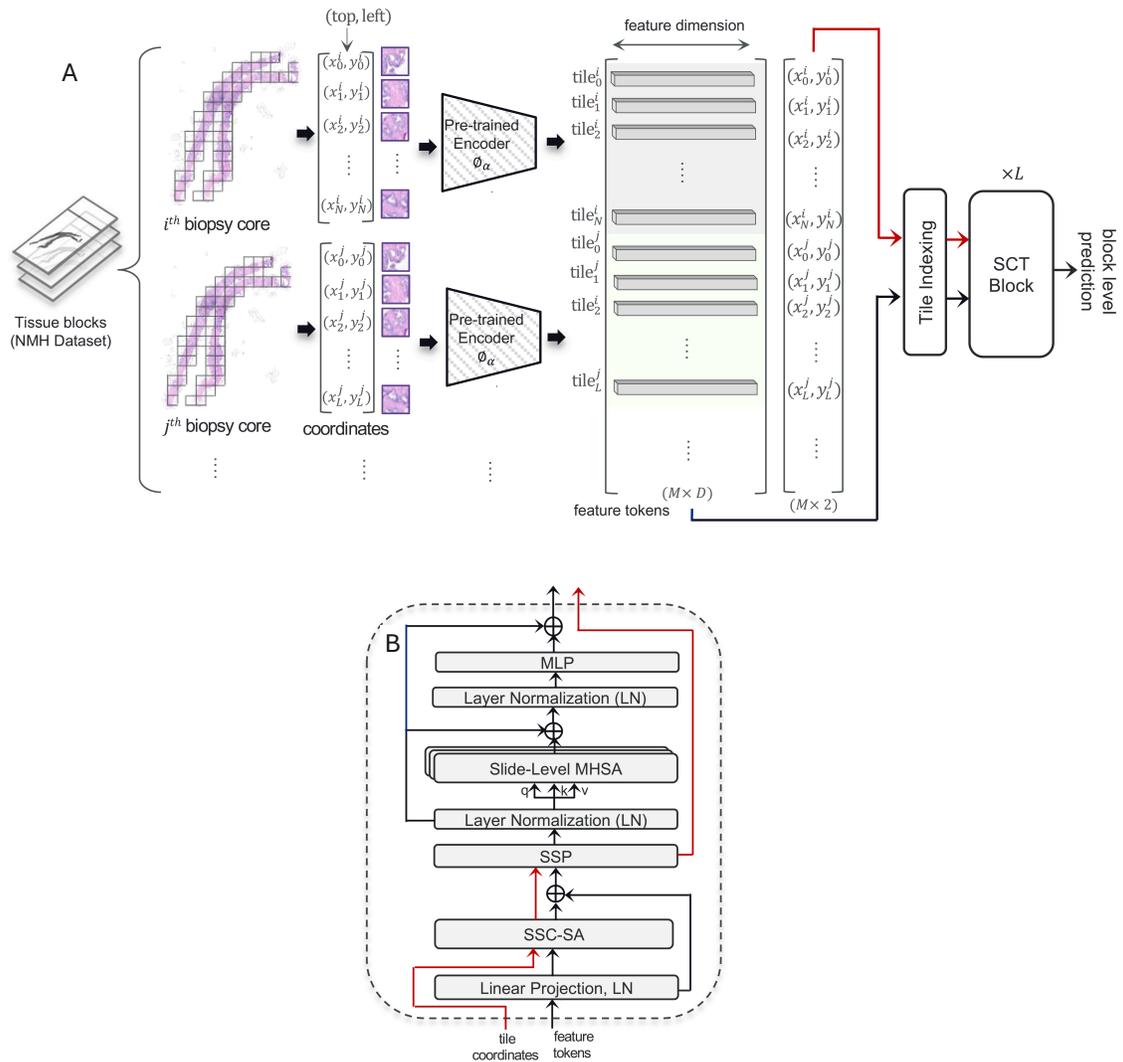

**Figure S7. SCT models for block-level prediction.** The detection and grading models built using SCT blocks that combine the efficiency of convolutional operations with a novel self-attention method to learn localized spatial and channel-wise relationships. (A) Block-level processing using SCT blocks. (B) Sparse Convolution Transformer (SCT) block.

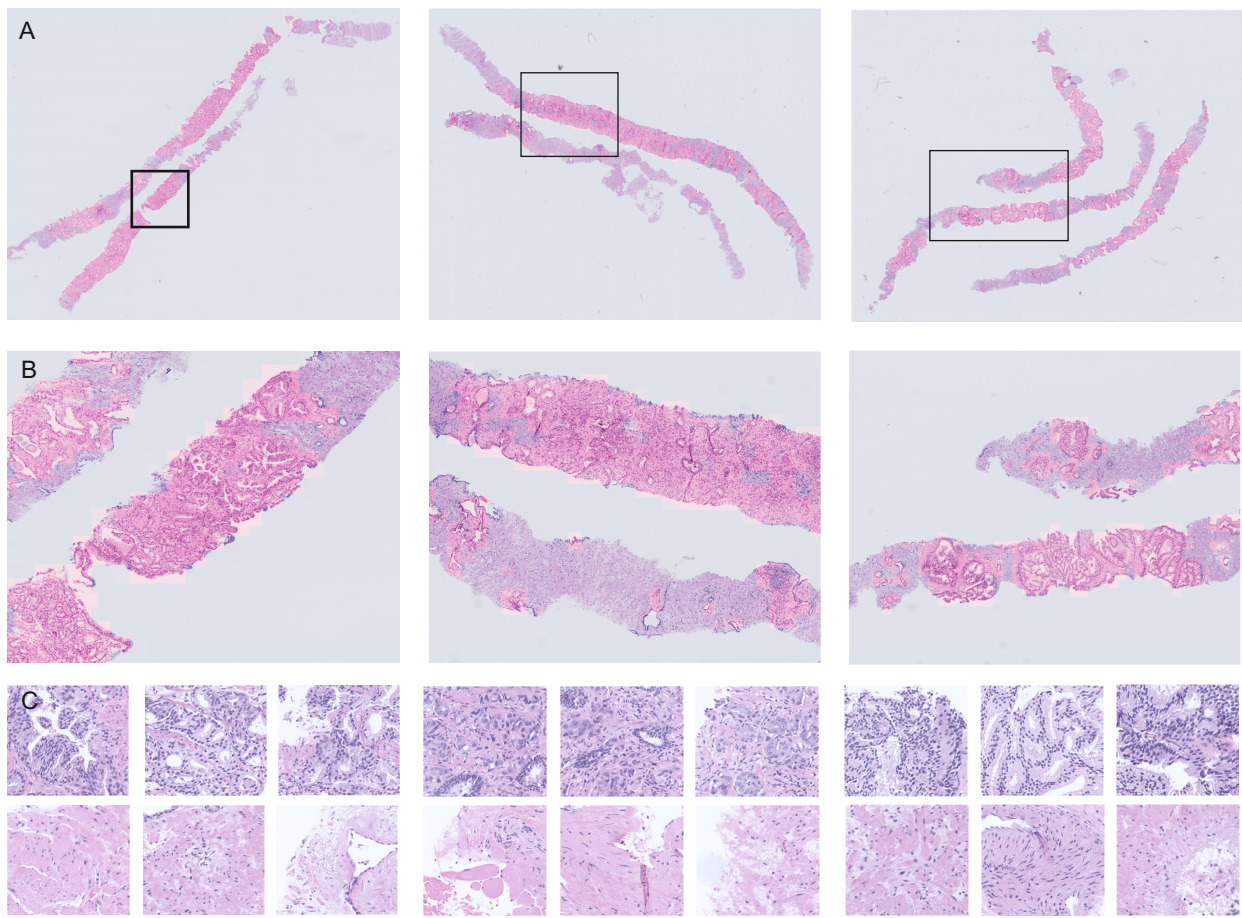

**Figure S8. Delineation of carcinoma for four sample whole slide images of NMH testing dataset.** (A) overlay of carcinoma on three whole slides. (B) Zoomed-in regions of interest (ROIs). (C) Corresponding carcinoma and benign tiles.

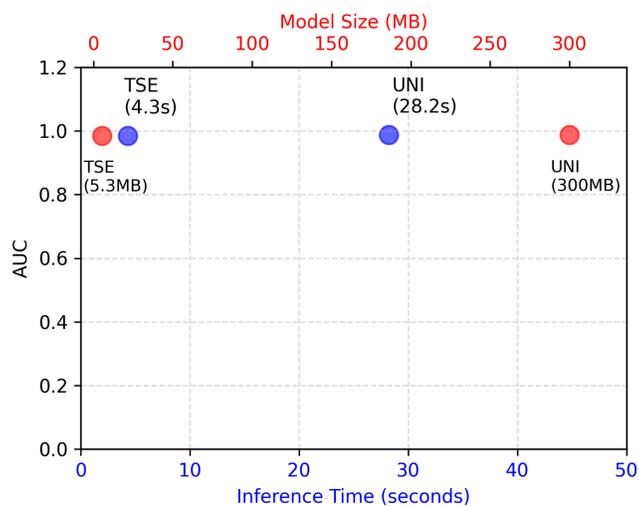

**Figure S9. Pareto plot comparing inference time, parameter count, and AUC for the UNI and TSE encoders.** The TSE exhibits approximately 50 times fewer parameters and achieves four times faster inference while maintaining comparable cancer detection performance (UNI: 98.7 AUC, TSE: 98.5 AUC).

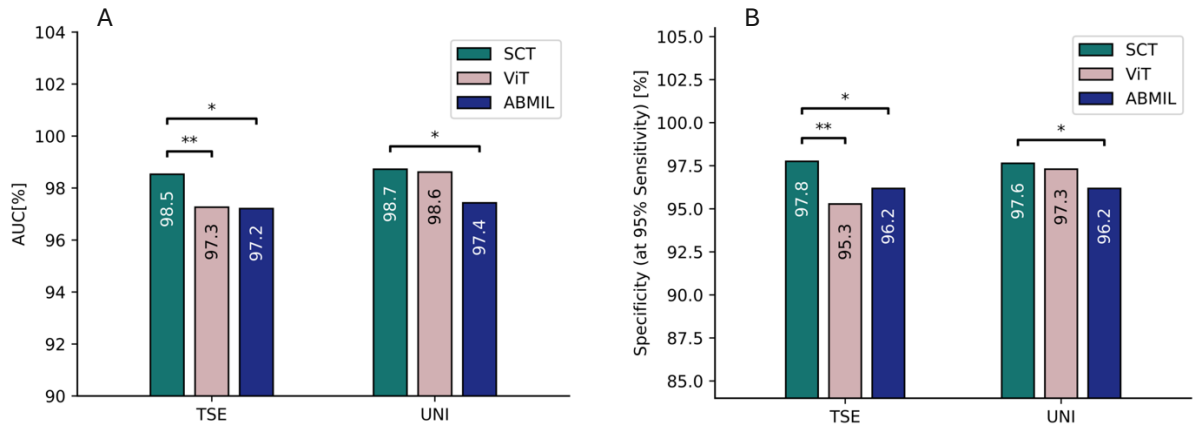

**Figure S10. Cancer detection performance comparison.** DeLong's test was used to determine the statistical significance, where significance levels were denoted by asterisks * for p < 0.05, ** for 0.01, and *** for p < 0.001. (A) AUCs obtained by SCT, ViT and ABMIL models with TSE and UNI. (B) Specificity at 95% sensitivity achieved by the models with the task-specific and UNI encoders.

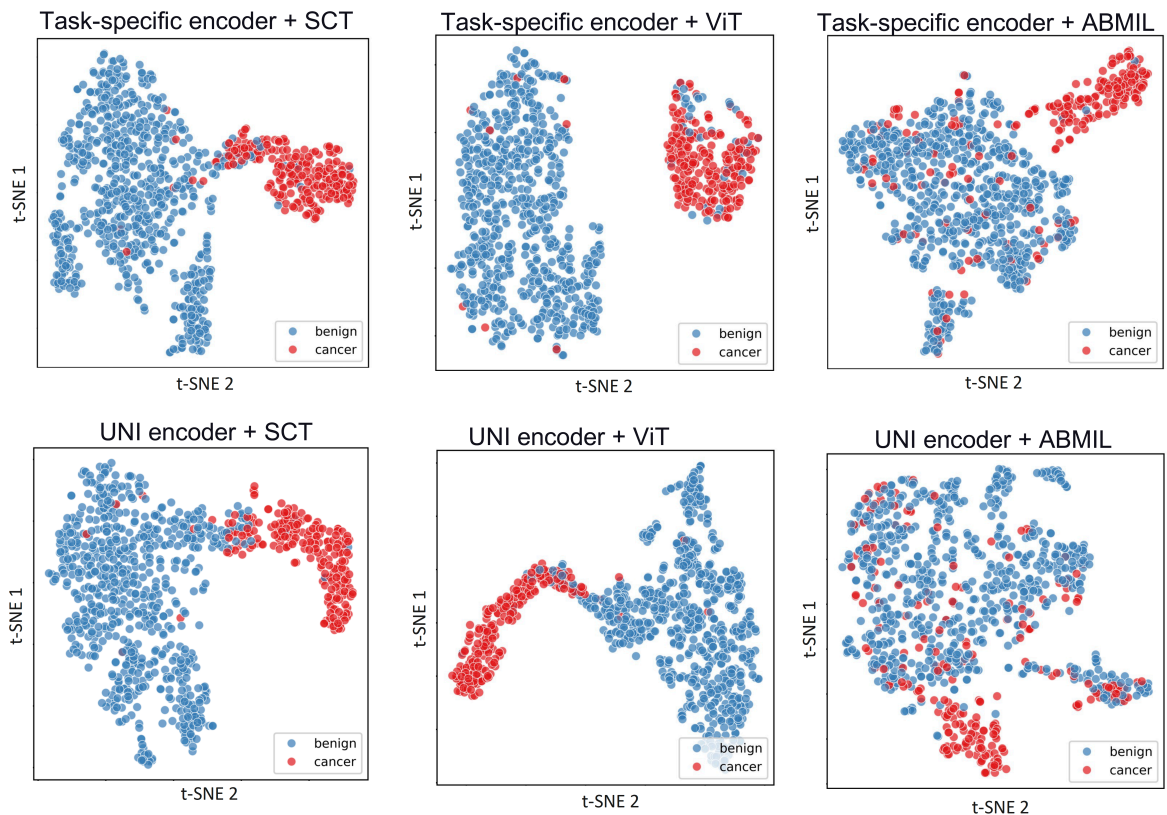

**Figure S11. T-SNE visualization of class separation from block-level feature embeddings learned by various encoders and models for cancer detection.**

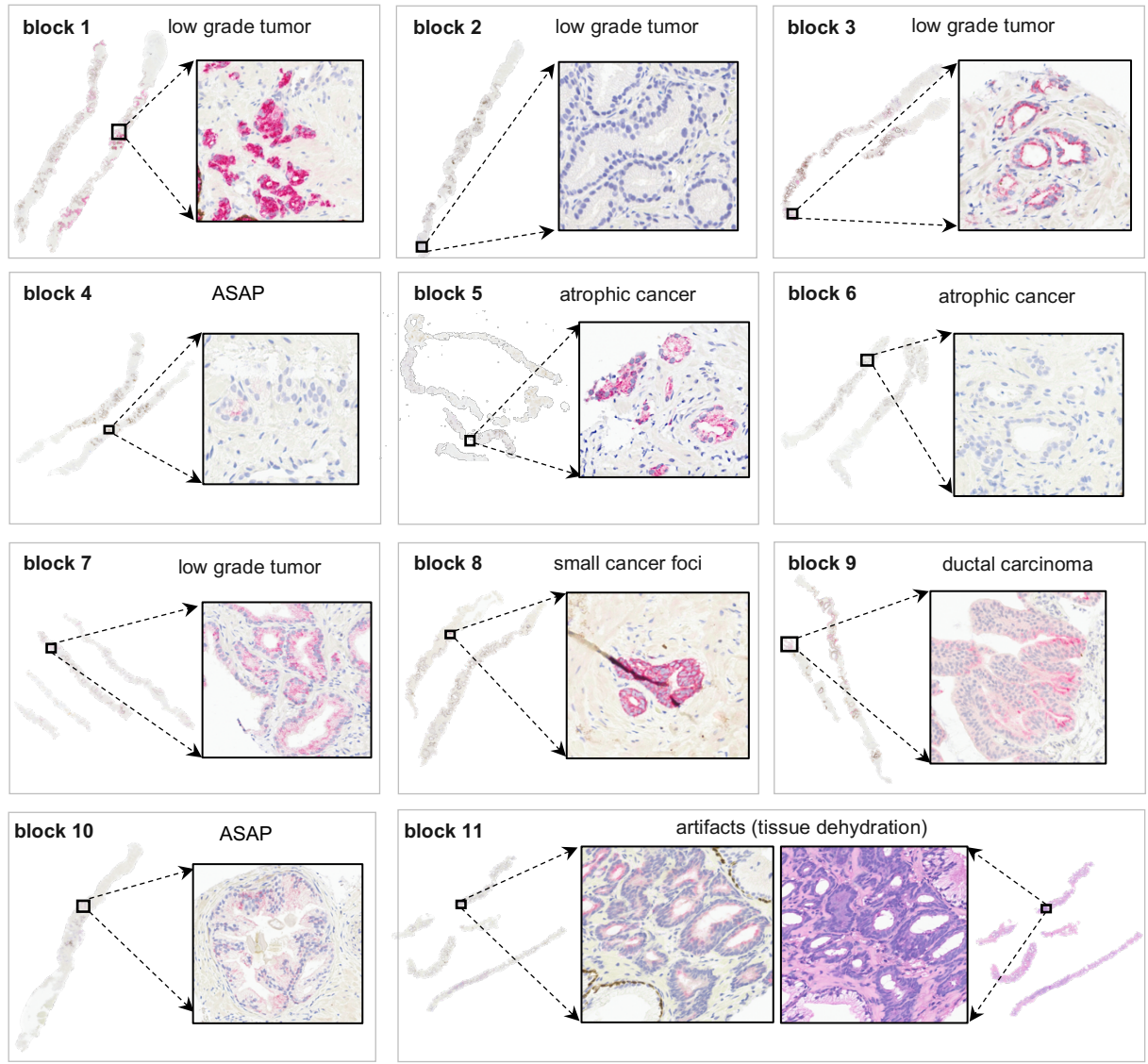

**Figure S12. Representative regions of interest from consensus review blocks.** Blocks 1, 2, and 3, contain low-grade Gleason 3+3 prostate cancer that was correctly detected by our model. Blocks 1 and 3 originate from patients who were initially diagnosed with ASAP. Block 2 was classified as benign, but originates from a biopsy with another positive core, so the patient diagnosis would be unchanged. Blocks 5-11, misclassified as benign by the model, were confirmed as cancer in the consensus review, consistent with the initial pathologist diagnoses. Of these, some blocks contain specific types of prostate cancer, such as atrophic cancer, ductal carcinoma and cribriform growth patterns, as well as tissue dehydration artifacts that may have contributed to the model's misdiagnosis. Our models operate on digital images H&E slides and IHC is shown here solely for visualization purposes.

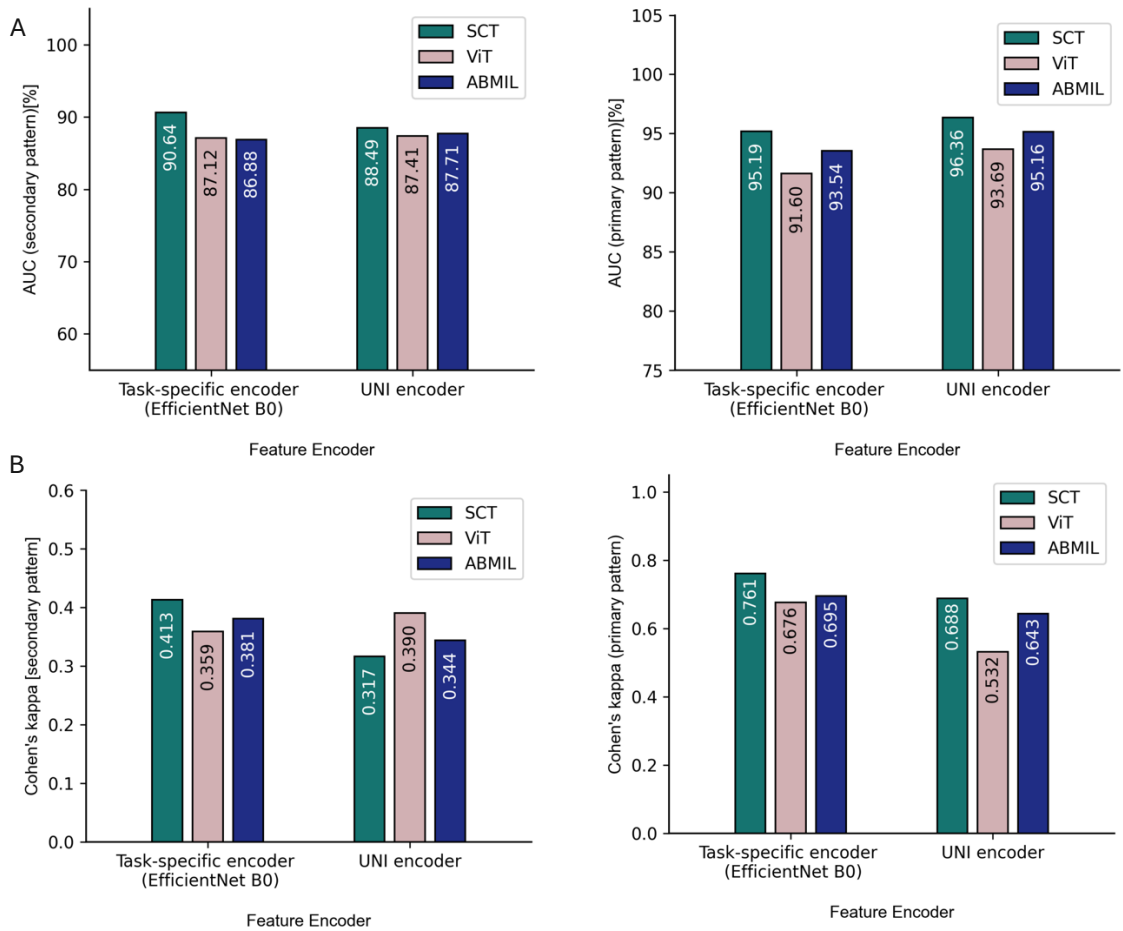

**Figure S13. Gleason grading performance comparison.** (A) AUCs of primary and secondary patterns prediction obtained by SCT, ViT and ABMIL models with the task-specific and UNI encoders. (B) Quadratic Cohen's kappa values of primary and secondary patterns achieved by the models with the task-specific and UNI encoders.

**Table S1. Reported performance of commercial and research tools in prostate cancer detection.**

|  | Ref. | Resolution | N | Source | Sensitivity (%) | Specificity (%) | AUC |
|---|---|---|---|---|---|---|---|
| Paige | [2] | slide | 232 | external | 96 | 98 | - |
|  | [3] | slide | 1876 | external | 97.7 | 99.3 | - |
|  | [8] | core | 600 | external | 98.9 | 93.3 | - |
|  |  | patient | 100 | external | 100 | 78.0 | - |
|  | [1] | slide | 113 | external | 98.8 | 87.9 | - |
| IBEX | [4] | slide | 2501 | internal | 99.6 | 90.1 | 99.7 |
|  |  |  | 1627 | external | 98.5 | 97.3 | 99.1 |
| Radboud | [5] | core | 535 | internal | 95.4 | 95.2 | 99.0 |
|  |  |  | 886 | external | 97.4 | 100 | 99.0 |
| Karolinska** | [6] | core | 1631 | internal | 94.9 | 99.0 | 99.7 |
|  |  |  | 330 | external | 97.7 | 89.8 | 98.6 |
| PANDA* | [7] | slide | 545 | internal | 98.1–99.7 | 91.9–96.7 | - |
|  |  |  | 741 | external | 97.6–99.3 (US) | 66.8–80.0 (US) | - |
|  |  |  | 330 | external | 96.2–99.2 (EU) | 70.5–87.9 (EU) | - |
| Northwestern |  | block | 890 | internal | 95.0 (TSE) | 97.76 (TSE) | 98.5 (TSE) |

*95% confidence interval for 15 tested models.
** Other operating points are available for sensitivity and specificity.

**Table S2. Reported performance of commercial and research tools in prostate cancer grading.**

| | Ref. | Resolution | N | Source | Metric | Value | Notes |
|---|---|---|---|---|---|---|---|
| Paige | [1] | slide | 113 | external | Sensitivity (%) | 100 | Benign, ISUP Grade 1-2 vs. ISUP Grade 3-5 |
| | | | | | Specificity (%) | 71.4 | |
| | | | | | κ (quadratic) | 0.86 | ISUP Grade |
| IBEX | [4] | slide | 1627 | external | Sensitivity (%) | 85.9 | Gleason score <7 vs. 7-10 |
| | | | | | Specificity (%) | 90.4 | |
| | | | | | AUC | 94.1 | |
| | | | | | Sensitivity (%) | 85 | Gleason pattern 3 or 4 vs. 5 |
| | | | | | Specificity (%) | 90.8 | |
| | | | | | AUC | 97.1 | |
| Radboud | [5] | core | 535 | internal | Sensitivity (%) | 91.8 | Grade group 1-2 vs. 3-5 |
| | | | | | Specificity (%) | 93.6 | |
| | | | | | AUC | 97.4 | |
| | | | 886 | external | Sensitivity (%) | 76.1 | Grade group 1-2 vs. 3-5 |
| | | | | | Specificity (%) | 78.3 | |
| | | | | | AUC | 85.5 | |
| | | | | | κ (quadratic) | 0.723 or 0.707 | Gleason score |
| Karolinska | [6] | core | 330 | external | κ (linear) | 0.62 | ISUP Grade |
| | | | | | | 0.7 | |
| PANDA | [7] | slide | 545 | internal | κ (quadratic) | 0.918-0.944 | ISUP Grade |
| | | | 330 | external | | 0.835-0.900 (EU) | |
| | | | | | | 0.840-0.884 (US) | |
| Northwestern | | block | 890 | internal | Sensitivity (%) | 96.6 (TSE) | Grade group 1-2 vs. 3-5 |
| | | | | | Specificity (%) | 94.9 (TSE) | |
| | | | | | AUC | 97.46 (TSE) | |
| | | | | | AUC | 98.07 (TSE) | ISUP Grade |
| | | | | | κ (quadratic) | 0.869 (TSE) | |

**Table S3. NMH cohort, including internal training, validation, and prospective testing sets.**

| | NMH Internal Training Set | NMH Internal Validation Set | Prospective Test Set |
|---|---|---|---|
| Slides | 18,990 | 2,406 | 1,737 |
| Blocks | 9,922 | 1,258 | 890 |
| Specimens | 1,019 | 128 | 95 |
| Subjects | 1,019 | 128 | 95 |

## Table S4. Overview of the PANDA dataset sources.

| Site | Dataset Size | Data Source | Staining | Scanning | Class Distribution | Training Label Source | Annotations |
|---|---|---|---|---|---|---|---|
| Radboud University Medical Center | 5160 WSIs | Pathology archives with a pathologist report between 2012 and 2017. | H&E | 3DHistech Panoramic Flash II 250 scanner | GG0: 967<br>GG1: 852<br>GG2: 675<br>GG3: 925<br>GG4: 768<br>GG5: 973 | semi-automatic labelling technique | Performed by trained students. |
| Karolinska Institute | 5456 WSIs | STHLM3 study with participants from the Stockholm County, Sweden, during the years 2013-2015. | H&E | Hamamatsu C9600-12 scanner (Hamamatsu Photonics, Hamamatsu, Japan) and an Aperio ScanScope AT2 scanner (Leica Biosystems, Wetzlar, Germany). | GG0: 1925<br>GG1: 1814<br>GG2: 668<br>GG3: 317<br>GG4: 481<br>GG5: 251 | Pathologists' annotations. | Performed by single experienced pathologist following routine clinical workflow. |

## Table S5. Architecture of SCT model.

| Model | Layers | SCT |
|---|---|---|
| Stage 1 | Linear projection | linear, 64 |
| | SSC-SA, SSP<br>Multi-head attention<br>MLP | $\begin{bmatrix} C:64, & k:3\times 3 \\ p:3, s:3, n_h:4 \end{bmatrix} \times 1$<br>gelu, 128 |
| Stage 2 | Linear projection | linear, 64 |
| | SSC-SA, SSP<br>Multi-head attention<br>MLP | $\begin{bmatrix} C:64, & k:3\times 3 \\ p:3, s:3, n_h:4 \end{bmatrix} \times 1$<br>gelu, 128 |
| Stage 3 | Linear projection | linear, 128 |
| | SSC-SA, SSP<br>Multi-head attention<br>MLP | $\begin{bmatrix} C:128, & k:3\times 3 \\ p:3, s:3, n_h:4 \end{bmatrix} \times 1$<br>gelu, 128 |
| | Linear projection | linear, 128 |
| Stage 4 | SSC-SA, SSP<br>Multi-head attention<br>MLP | $\begin{bmatrix} C:128, & k:3\times 3 \\ p:3, s:3, n_h:4 \end{bmatrix} \times 1$<br>gelu, 128 |
| | Parameters | 1.38 M |
| | Flops | 863694 |

## Table S6. Performance comparison of the SCT, ViT, and ABMIL models with UNI and task-specific encoders for cancer detection.

| Model | SCT | | ABMIL | | ViT | |
|---|---|---|---|---|---|---|
| Encoder | Task-spec. | Uni | Task-spec. | Uni | Task-spec. | Uni |
| True positives (TPR) | 209 (95.0) | 211 (95.9) | 209 (95.0) | 204 (92.7) | 211 (95.9) | 205 (93.2) |
| True negatives (TNR) | 655 (97.8) | 653 (97.5) | 642 (95.8) | 646 (96.4) | 637 (95.0) | 653 (97.5) |
| False positives (FPR) | 15 (2.2) | 17 (2.5) | 28 (4.2) | 24 (3.6) | 33 (5.0) | 17 (2.5) |
| False negatives (FNR) | 11 (5.0) | 9 (4.0) | 9 (4.0) | 16 (7.2) | 9 (4.0) | 15 (6.8) |

**Table S7. Summary of consensus review findings for IHC-requested biopsy blocks**

| Model | Block | Initial diagnosis | Consensus review | Consensus remarks |
|---|---|---|---|---|
| **cancer** | 1 | ASAP, PINATYP | cancer | low-grade low-volume tumor (GG1) |
| | 2 | benign | cancer | low-grade low-volume tumor (GG1). This specimen contained another GG1 block and so this discrepancy would not impact diagnosis. |
| | 3 | ASAP | cancer | low-grade low-volume tumor (GG1) |
| **benign** | 4 | cancer | benign | atrophic with HGPIN/ASAP; cancerous glands present but insufficient for a cancer diagnosis |
| | 5 | cancer | cancer | atrophic cancer |
| | 6 | cancer | cancer | atrophic cancer from HE, but not evident in IHC stain |
| | 7 | cancer | cancer | cribriform, straight nuclear border, nucleoli |
| | 8 | cancer | cancer | tiny cancer foci with atypical morphology |
| | 9 | cancer | cancer | ductal cancer |
| | 10 | cancer | cancer | atypical small acinar proliferation, nuclear atypia, luminal crystalloid structures |
| | 11 | cancer | cancer | tissue dehydration artifact |